\documentclass[12pt]{article}
\usepackage{amsmath, amsfonts, amssymb, amsthm, graphicx, enumerate, setspace, color, colortbl, multirow, verbatim, array, float, longtable, booktabs, tablefootnote, plain, dsfont, color, makecell}

\usepackage[margin=1in]{geometry}
\usepackage[labelfont=bf]{caption}
\usepackage[title]{appendix}

\usepackage{url}
\usepackage[breaklinks]{hyperref}
\hypersetup{
    colorlinks = true,
    linkcolor = {blue},
    citecolor = blue,
    urlcolor = {blue}
}

\usepackage{natbib}
 \bibpunct[, ]{(}{)}{,}{a}{}{,}%
\DeclareTextFontCommand{\emph}{\em}

\usepackage{titlesec}

\titlespacing\section{0pt}{10pt plus 4pt minus 2pt}{0pt plus 2pt minus 2pt}
\titlespacing\subsection{0pt}{10pt plus 4pt minus 2pt}{0pt plus 2pt minus 2pt}
\titlespacing\subsubsection{0pt}{10pt plus 4pt minus 2pt}{0pt plus 2pt minus 2pt}

\usepackage{pstricks}
\newrgbcolor{monkey}{1 0.0 0.0}
\newrgbcolor{gorilla}{0.0 0.0 0.75}

\begin{document}
\begin{titlepage}

\vspace{-1cm}

\title{\vspace{-1cm}The Effect of Product Recommendations on Online Investor Behaviors}

\author{Ruiqi Rich Zhu\footnote{Scheller College of Business, Georgia Institute of Technology, Atlanta, Georgia 30308. E-mail: $\textrm{rich.zhu@scheller.gatech.edu}.$},
$\ $ Cheng He\footnote{Wisconsin School of Business, University of Wisconsin-Madison, Madison, Wisconsin 53706. E-mail: $\textrm{cheng.he@wisc.edu}.$},
$\ $ Yu Jeffrey Hu\footnote{Mitchell E. Daniels, Jr. School of Business, Purdue University, West Lafayette, Indiana 47907. E-mail: $\textrm{yuhu@purdue.edu}.$}}

\vspace{-1.5cm}
\date{}
\maketitle
\vspace{-1cm}
\thispagestyle{empty}%


\begin{abstract}
\noindent Despite the popularity of product recommendations on online investment platforms, few studies have explored their impact on investor behaviors. Using data from a global e-commerce platform, we apply regression discontinuity design to causally examine the effects of product recommendations on online investors' mutual fund investments. Our findings indicate that recommended funds experience a significant rise in purchases, especially among low socioeconomic status investors who are most influenced by these recommendations. However, investors tend to suffer significantly worse investment returns after purchasing recommended funds, and this negative impact is also most significant for investors with low socioeconomic status. To explain this disparity, we find investors tend to gather less information and expend reduced effort in fund research when buying recommended funds. Furthermore, investors' redemption timing of recommended funds is less optimal than non-recommended funds. We also find that recommended funds experience a larger return reversal than non-recommended funds. In conclusion, product recommendations make investors behave more irrationally and these negative consequences are most significant for investors with low socioeconomic status, which can amplify wealth inequality among investors in financial markets.
\\ \\

\noindent \emph{Keywords:} Fintech, Product Recommendations, Behavioral Finance, Mutual Fund
\end{abstract}
\end{titlepage}

\newpage
\setstretch{1.6}

\section{Introduction}

The number of online retail investors has surged. In 2020, U.S. online investors bought an average of \$1 billion in listed equities daily, marking a fivefold rise from pre-2020 figures \citep{retailinvest}. By Q3 2021, equity market retail flow had nearly doubled since 2010 \citep{creditsuisse}. Online investment platforms like Robinhood and SoFi have also witnessed tremendous growth \citep{fitzgerald}. These platforms, emulating e-commerce sites, use features like product recommendations to help investors discover good financial products and achieve high investment returns. However, in countries like China and the U.S., regulations prevent them from offering algorithm-driven personalized stock or fund recommendations unless provided by a registered financial advisor. Therefore, many platforms adopt general recommendations by highlighting products with strong past performance for all investors, akin to e-commerce sites showcasing popular and highly-rated products.

For platforms selling consumer goods, literature \citep{de2010technology, brynjolfsson2011goodbye, hosanagar2014will, goldfarb2019digital} has shown that product recommendations can boost sales for the platform by highlighting products' ratings and sales \citep{adomavicius2005toward}, therefore offering strong quality signals to consumers \citep{pathak2010empirical}. However, for financial products, the efficacy of recommendations is uncertain. Studies have demonstrated that historical returns do not reliably predict future returns, so rational investors should not treat them as reliable quality signals when purchasing financial products \citep{kahn1995does, clifford2021salience, barber2021attention}. Thus, while product recommendations effectively drive consumer goods sales by providing reliable quality signals, their influence on investor behaviors and financial product sales, when based on unreliable quality signals, remains unclear.

Furthermore, consumers typically have a similar perception of quality signals for consumer goods  \citep{zhao2013modeling}. In contrast, financial products elicit varied perceptions of quality signals, depending on an investor's knowledge and experience \citep{havakhor2021tech}. Unsophisticated investors, for instance, often chase financial products with high historical returns and tend to suffer losses due to trend-chasing \citep{barber2021attention}. With this study, we investigate specifically whether unsophisticated investors might be more likely to adopt an online investment platform's recommendations, as well as whether they suffer diminished outcomes after purchasing the recommended products. Recommender systems ideally should help reduce knowledge gaps among consumers, by providing a reliable quality signal to less-educated consumers \citep{oestreicher2012recommendation}; we consider instead whether, in financial markets, recommender systems actually might amplify wealth inequality among investors, as a dark-side implication.
 
To address these possibilities, we collect a novel data set from a leading global e-commerce platform that sells financial products directly to investors.\footnote{The platform ranks among the top five global e-commerce companies, according to GMV \citep{ecommercerank}.} This unique data set includes investor-level mutual fund purchases, redemption, and investor profiles, rare in prior literature.\footnote{Most studies of mutual funds focus on market-level mutual fund inflows and outflows, without considering information about individual investors who purchase the fund.} The platform recommends top-performing funds based on annual returns, a practice adopted by other prominent platforms like Alipay, Tenpay, and TD Ameritrade. When investors view a fund, the platform recommends the top five funds in the same fund type as the viewed fund with the highest annual returns on the viewing date. Using a sharp regression discontinuity design (RDD), we compare sales of funds that just reach the annual return cutoff to be recommended with non-recommended funds that narrowly miss this cutoff. Using this approach, we can look for discontinuous jumps in fund purchases that follow discontinuous changes in the fund's recommendation status. Our findings show that on average, being featured as a recommended fund increases purchases by 5.6\%. Therefore, product recommendations affect investors' behaviors, even though they are not reliable quality signals from a rational investor's perspective. 

Next, we investigate how investor characteristics impact the effect of product recommendations. Notably, investors with lower socioeconomic status, (lower education level, lower income level, and higher liabilities) are more influenced by recommendations. In follow-up analyses, we calculate profits (losses) experienced by investors when they purchase recommended versus non-recommended funds using their purchase and redemption histories. Our analysis reveals that investors tend to suffer significantly worse performance after purchasing recommended funds. This worse performance is also most salient for investors with low socioeconomic status. Thus, investors of low socioeconomic status, who arguably should be better protected in financial markets, instead tend to have worse investment outcomes when they follow product recommendations provided by online investment platforms. This might amplify wealth inequality between investors with varying socioeconomic statuses. 

Finally, we investigate why purchasing recommended funds versus non-recommended ones is associated with worse investment returns. First, we observe that while investors' fund consideration set size does not change, there is a significant reduction in search time spent on researching funds when purchasing recommended funds, leading to less informed investment decisions. This phenomenon is most pronounced among low socioeconomic status investors, who are also most responsive to product recommendations. Probing further into the kind of funds that receive diminished search time by investors, we find when investors purchase recommended funds (versus non-recommended ones), they significantly reduce the time allocated to exploring similar alternative funds that share the same fund type as the purchased funds. However, this reduction does not extend to funds with different fund types from the purchased fund. Additionally, our simulation analysis further indicates that investors who purchase recommended funds tend to redeem at a worse time than those who purchase non-recommended funds. Surprisingly, this lapse in redemption timing is predominantly observed among low socioeconomic status investors. Lastly, upon accessing fund future performance, data indicates that recommended funds generally underperform their non-recommended counterparts. Potential explanations for this underperformance may include regression to the mean and initial price overreaction \citep{de1985does, atkins1990price, huang2010return}. 

Our study makes several contributions. First, we expand the existing body of literature on the economic impacts of product recommendations. While prior research has emphasized that product recommendations can boost sales by offering reliable quality signals \citep{pathak2010empirical}, our study highlights their influence on investor behaviors even in situations where the quality signals are not reliable. Second, we contribute to the literature on the potential dark side of recommendations by demonstrating how purchasing recommended funds (versus non-recommended funds) is associated with worse investment returns. This disparity in returns is particularly pronounced among investors with low socioeconomic status who also tend to rely the most on these recommendations. This exposes a disconcerting facet of product recommendations in the realm of finance, indicating their potential to encourage irrational behaviors and exacerbate economic inequality \citep{adomavicius2013recommender, walsh2020algorithms, zhang2021welfare}. 

Third, our study enriches the field of behavioral finance by utilizing extensive individual-level data to investigate the influence of features like product recommendations on investors' purchasing decisions and investment returns. Furthermore, we delve into subsequent behavioral changes and potential explanations for worse performance when investors follow recommendations by examining their behaviors in the stages of purchase and redemption. When following product recommendations, investors spend significantly less search time researching funds and also make less favorable choices when it comes to redeeming their investments. These findings contribute significantly to the behavioral finance literature, shedding light on the adverse effects on investor behavior that can arise when technology makes financial information more accessible and prominent to investors. \citep{barber2002online, clifford2021salience, havakhor2021tech}.

\section{\label{literature section}Literature Review}
In reviewing existing literature related to the study of product recommendations for financial products, we consider two main streams, pertaining to (1) the economic effects of product recommendations and (2) behavioral finance.

\subsection{\label{economic recommend}Economic Effects of Product Recommendations}
Existing literature on the economic effects of product recommendations suggests that they increase firm sales and revenues \citep{de2010technology, pathak2010empirical, brynjolfsson2011goodbye, hosanagar2014will, kawaguchi2019effectiveness, lee2021product}. \cite{de2010technology} show that product recommendations can increase sales of both promoted and non-promoted products. \cite{pathak2010empirical} point out that the strength of the recommendation is relevant to the positive effect on sales. Product recommendations also have a volume effect that encourages consumer consumption and increases both views and final conversion rates \citep{hosanagar2014will, lee2021product}. \cite{ursu2018power} find recommendations to decrease search costs and increase the probability of a match with a seller. Paper by \cite{lee2021product} provide a detailed literature review on the economic effects of product recommendations.

Whereas prior work focuses on the effect of product recommendations for consumer goods such as books and clothes, we extend prior research to financial products, which are distinct from consumer goods. Though product recommendations serve as reliable quality signals for consumer goods using past ratings and sales, their reliability is questionable for financial products due to their reliance on historical returns. The value of financial products lies in unobserved future returns, which cannot be reliably predicted using historical returns \citep{kahn1995does, jain2000truth, choi2020carhart}. Thus, the relationship between recommendations and increased sales is uncertain for financial products. Our study aims to fill this gap by analyzing how product recommendations influence online investor behaviors.

Furthermore, product recommendations can unintentionally skew consumer judgment and choices when purchasing consumer goods. \cite{adomavicius2013recommender} suggest that product recommendation ratings can shape consumers' preferences, often causing them to place undue importance of these suggestions. Even when presented with random recommendations, consumers' perceived value of a product can significantly shift \citep{adomavicius2018effects}. Moreover, consumers are found to be less sensitive to price for prominently recommended products, which might reduce their overall benefits from the purchase \citep{zhang2021welfare}. Additionally, there's an over-reliance on algorithm-generated recommendations, even if they are inferior \citep{banker2019algorithm}. Our study extends this literature on the potential pitfalls of product recommendations, specifically in the context of financial product investments. We also identify several negative impacts of product recommendations that make investors behave more irrationally and realize worse investment outcomes. As our research centers on the economic effect of product recommendations, we do not discuss the rich literature that studies the technical design of personalized recommendations \citep{shen2018behavior, tong2020personalized, ke2022information}. For a detailed literature review on the taxonomy and technical details of recommender systems, please refer to the literature review by \cite{adomavicius2005toward} and \cite{roy2022systematic}.

\subsection{\label{behavioral finance}Behavioral Finance}
Behavioral finance literature highlights factors that shape investment decisions. Investors often opt for financial products that stand out due to exceptional performance \citep{barber2008all}, marketing efforts \citep{barber2005out}, customer satisfaction rating \citep{sorescu2016customer}, product promotion \citep{jiang2021consumer}, herding by peers \citep{zhang2012rational}, or visible fees \citep{barber2005out}. For instance, studies show that investors gravitate toward stocks or mutual funds that are currently in the media spotlight, indicating that media presence reduces the effort investors have to make when choosing where to invest \citep{sirri1998costly, tetlock2007giving}. Additionally, investments with notably volatile returns tend to attract more attention \citep{clifford2021salience}. Our research delves deeper, exploring the impact of automated product recommendations on online platforms. These recommendations not only shift investor attention but also continuously update what products are recommended. We utilize detailed investor-level data, allowing us to analyze recommendations' effects on individual investors. This in-depth approach helps us understand how different types of investors respond and offers insights into their decision-making processes and investment outcomes, areas not extensively covered in previous studies.

Furthermore, behavioral finance studies explore how investor characteristics impact their investment decisions. Studies by \cite{barber2001boys} suggest men, driven by overconfidence, trade more often than women and reduce their net gains. Cognitive abilities also play a role: high-IQ investors tend to avoid funds with high management fees \citep{grinblatt2016iq}. Furthermore, investors who are younger, less educated, or have lower incomes often allocate a bigger chunk of their wealth to individual stocks, leading to frequent trading and suboptimal results \citep{anderson2013trading}. Utilizing our rich data, we categorize investors by socioeconomic status through their demographic information. Our findings build upon previous research, highlighting that investors of low socioeconomic status are most influenced by product recommendations, often to their detriment. These investors often invest based on recommendations without comprehensive research, resulting in significant investment losses.

Our work is also related to research that studies how expert recommendations influence fund flows. A study by \cite{cookson2021best} examines expert-curated mutual fund suggestions on investment platforms. Given that these expert recommendations are often viewed as reliable quality signals, they noticeably boost sales. In contrast, the recommendations in our study are purely based on past performance and lack human intervention, making them less reliable signals for investors. Thus, the impact of such automated product recommendations is unclear. While \cite{cookson2021best} illustrate the link between expert advice and sales, our method employs a regression discontinuity design to understand the direct causal effect of these automated recommendations on investor behaviors in investment platforms.

Another relevant work by \cite{barber2021attention} highlights how Robinhood app’s unique ``Top Mover'' list drives investors to herd and buy attention-grabbing stocks. As the “Top Mover” list displays stocks with notable price fluctuations, it sends a different signal from the product recommendations in our study, which rely on historical returns. While this paper emphasizes the impact of app features on sales using broad order flow data, our research delves deeper, examining the granular effects of recommendations on individual investors. Our rich dataset, encompassing both purchase and redemption histories, allows us to study both purchasing behaviors and individual performance outcomes. This granularity facilitates exploration into how product recommendations might differentially influence investor return outcomes across socioeconomic statuses. Additionally, by leveraging click stream data, we scrutinize the intricacies of investors' fund-searching and decision-making processes when purchasing funds.

\section{Institutional Background and Data}
\subsection{\label{background}Recommendation in Online Investment Platforms}
The surge in online retail investor participation has led investment platforms to borrow features such as product recommendations from traditional e-commerce. They employ these features to improve investors' experience with the goal of simplifying investors' decision-making process, helping investors discover promising financial products, and empowering them to achieve attractive investment returns.\footnote{\url{https://www.investopedia.com/best-online-brokers-4587872}} However, investment platforms cannot provide personalized recommendations to investors the same way as e-commerce platforms do.\footnote{\url{https://srinstitute.utoronto.ca/news/the-art-and-science-of-recommender-systems-insights-from-spotify}} Personalized recommendations for financial products, like stocks or mutual funds, face stringent regulations on investment platforms. In the U.S., personalized stock and financial recommendations are generally restricted to professional investment advisers who have met the requirements set forth by regulations and laws. For instance, the Investment Advisers Act of 1940 requires that investment advisers register with the SEC and adhere to regulations designed to protect investors.\footnote{\url{https://www.investopedia.com/terms/i/investadvact.asp}} Several anti-fraud provisions also restrict the provision of personalized investment advice unless the provider is properly qualified and registered. This includes Sections 206(1) and 206(2) of the Investment Advisers Act and Section 10(b) and Rule 10b-5 of the Securities Exchange Act of 1934.\footnote{\url{https://www.sec.gov/about/about-securities-laws}} 

Like the U.S., China enforces strict regulations concerning financial product recommendations. The financial markets in China are regulated by various bodies, including the China Securities Regulatory Commission (CSRC), the People's Bank of China (PBOC), and the State Administration of Foreign Exchange (SAFE), among others. As stipulated by the Securities Law of the People's Republic of China, providing investment advice in relation to securities without proper qualifications or outside of approved institutional frameworks could lead to penalties.\footnote{\url{http://www.npc.gov.cn/englishnpc/}} Online platforms aren't exempt; unauthorized individuals providing financial advice can be held accountable for disseminating investment advice without a proper license. Similarly, regions like the European Union maintain stringent controls on financial product recommendations,\footnote{\url{https://finance.ec.europa.eu/regulation-and-supervision/financial-services-legislation/implementing-and-delegated-acts/markets-financial-instruments-directive-ii_en}} mirroring the regulatory standards set in both the U.S. and China.

Investment platforms, due to those regulatory constraints, often resort to ranking-based product recommendation designs that comply with financial regulations. Such designs showcase financial products, like mutual funds, with the highest historical returns to all investors. While not directly advising investors, highlighting products with high historical returns implicitly sends quality signals that those are ``good'' products. This indirect method effectively serves as product recommendations. Major investment platforms, including TD Ameritrade in the U.S. and Alipay and East Money in China, use this approach, recommending products with notable historical returns to investors.

\subsection{\label{data}Data and Empirical Context}
We collect a unique data set from a leading global e-commerce platform that operates on a business-to-consumer (B2C) model. The financial firms on the platform sell various products, such as mutual funds and insurance, as well as provide services like crowdfunding and loans directly to investors. In this study, we focus on mutual funds, one of the most popular financial products, to examine the impact of recommendations.

\subsubsection{\label{recommendations design}Product Recommendations Design}
\begin{figure}[ht]
    \centering
    \caption{Product Recommendation Illustration}
    \includegraphics[scale=0.5]{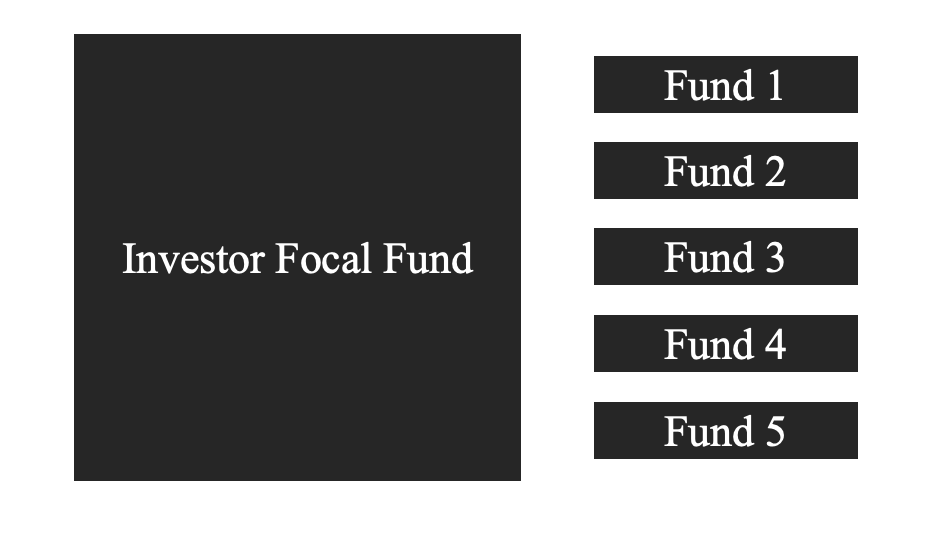}
    \label{fig:RSIllustration}
\end{figure}

Similar to many online investment platforms, this platform highlights mutual funds that have yielded high historical returns. When investors view a fund (referred to as ``the focal fund'' in our study), the platform suggests five other funds on the right side of the focal fund detail page. All these recommended funds belong to the same fund type as the focal fund and are listed in descending order based on their annual returns on the focal fund view date. Our data includes five types of funds: hybrid, stock, bond, money market, and QDII.\footnote{QDII (Qualified Domestic Institutional Investor) funds enable domestic investors to invest in securities in foreign markets. More details at \url{https://www.investopedia.com/terms/q/qdii.asp}} As depicted in Figure~\ref{fig:RSIllustration}, funds that rank fifth or higher in terms of annual returns are recommended. For example, if the focal fund is a hybrid fund, then the fourth fund displayed would represent the hybrid fund with the fourth highest annual return among all hybrid funds on the focal fund view date.

\subsubsection{\label{investor behavior}Investor Transactions}
Our data set comprises a random sample of 12,441 individual investors, each identified by a unique encrypted ID. We have their fund transaction histories spanning from January 2015 to December 2016. For each transaction, we note the date, fund code, investor ID, transaction type (purchase or redemption), and order amount.

\subsubsection{\label{fund attributes}Mutual Fund Attributes}
Our dataset encompasses comprehensive information on 2,999 funds available on the platform from January 2015 to December 2016. Details include the fund code, name, fund type, fund establish date, asset size, management fees, purchase fees, custodial fees, fund risk levels, and a dummy variable indicating valuable fund brand.\footnote{Determined by Easymoney, the most prominent fund research website in China} We also construct the variable fund tenure, as the difference between the fund transaction date and fund establish date, measured in days. The data also provides fund tenure and performance metrics including weekly, monthly, three-month, annual, and lifetime returns\footnote{Measure of how much the value of the fund has increased since the fund is established} on each day.

\subsubsection{\label{investor characteristics}Investor Characteristics}
We also obtain demographic information for each investor from a mandatory questionnaire when the investor creates the account, as summarized in Table~\ref{tab:investorSummary} in Appendix~\ref{A}. This information includes investors’ levels of income, education, and liabilities. In the following sections, we use this information to segment investors into different socioeconomic status groups. An in-depth discussion on this segmentation can be found in Section~\ref{investor level effect}. Our investor sample closely resembles the mutual fund investor demographics in China\footnote{\url{https://edu.efunds.com.cn/c/2022-01-14/497432.shtml}} though with a slightly higher percentage of low socioeconomic status investors. 

\section{\label{main effect}Effect of Product Recommendations on Fund Purchases}
In this section, we examine the effect of product recommendations on fund sales. A critical challenge in identifying the causal effects of product recommendations stems from the potential intrinsic differences between recommended and non-recommended funds. A direct comparison between the two might yield biased estimation results. To address this, we employ a regression discontinuity design (RDD) that leverages the specific design in which the platform recommended funds (details in Section~\ref{recommendations design}). We further validate our findings through a series of robustness checks.

\subsection{\label{RDD}Empirical Strategy: Regression Discontinuity Design}
\begin{figure}[!ht]
    \centering
    \caption{RDD Illustration}
    \includegraphics[scale=0.45]{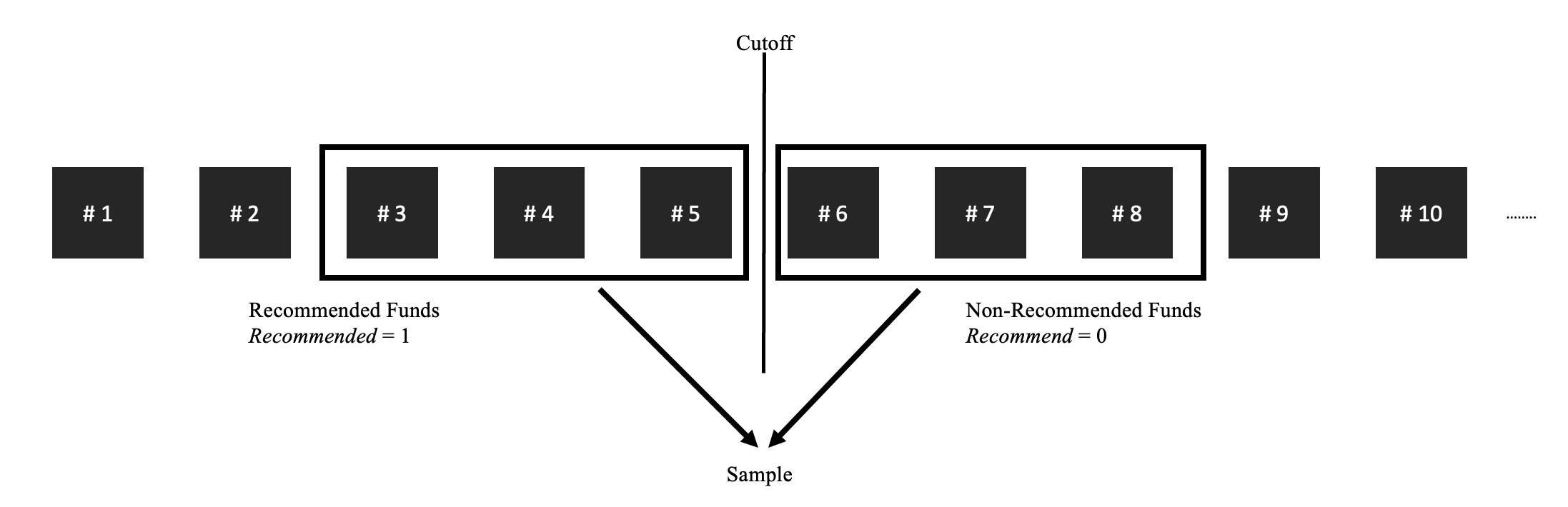}
    \label{fig:RDDesign}
\end{figure}
In our study, the top five performing funds based on annual returns are recommended to investors. We exploit this product recommendation design and narrow our focus to funds that rank third to fifth by annual return, i.e., funds that barely make the recommendation cutoff, and funds that rank sixth to eighth, i.e., funds that just fall short of this threshold. This bandwidth is optimal in mean squared error and funds within this window exhibit better covariate balance and overlap \citep{cattaneo2022regression}. To strengthen our findings, we also conduct a robustness check with a narrower bandwidth and find consistent results. As illustrated in Figure~\ref{fig:RDDesign}, funds ranking third to fifth get recommended to the investor; funds ranking sixth to eighth do not.

To formally evaluate the influence of product recommendations on fund purchases, we employ the following log-linear model for our analysis:

\begin{eqnarray} \label{eq:mainEquation}
\begin{aligned}
\log{(NumberOfPurchases_{it})}= &\beta\ast Recommended_{it}+ f(AnnualReturn_{it}) + \\
& \theta\ast X_{it}+\alpha_{i}+\gamma_{t}+\xi_{t}+\epsilon_{it}
\end{aligned}
\end{eqnarray}

Our unit of analysis is fund ($i$) – day ($t$) level. The variable $Recommended_{it}$ is a binary indicator set to 1 if fund $i$ ranks third, fourth, or fifth in annual returns among funds with the same type as fund $i$ on day $t$. Conversely, $Recommended_{it}$ is 0 if fund $i$ ranks sixth, seventh, or eighth in annual returns among funds with the same type as fund $i$ on day $t$.
log($NumberOfPurchases_{it}$) denotes the logarithm of the number of purchase transactions for fund $i$ on day $t$. The function $f(AnnualReturn_{it})$ represents a local polynomial function of the fund’s annual return, up to the degree of 2,\footnote{The choice of this degree is based on obtaining the highest BIC.} that allows for a smooth relationship between annual return and fund purchases around the cutoff. $\alpha_{i}$ corresponds to fund type fixed effects, which capture variations in sales among funds sharing the same fund type. $\gamma_{t}$ are month fixed effects, capturing seasonal trends and market fluctuations that impact sales of all funds.  $\xi_{t}$ are day-of-week fixed effects that incorporate variations in fund sales on different days in the week for all funds. Through $X_{it}$, we control fund attributes including management fees, purchase fees, custodial fees, tenure, a dummy for valuable fund brand, asset size, and one-week, three-month, six-month, and lifetime returns. These control variables capture potential impacts of fund attributes on the number of fund purchases. Finally, $\beta$ is our parameter of interest. Given the identifying assumption that other determinants of fund purchases are continuous at the cutoff and the running variable annual return cannot be manipulated, we use $\beta$ to capture product recommendations effects on fund purchases \citep{thistlethwaite1960regression, lee2010regression, goldfarb2022conducting, cattaneo2022regression}. 

\subsection{\label{identification check}Validity of Identification Assumptions}
Before we present the estimation results, we first test the validity of our RDD design by examining two key identification assumptions: the continuity of observable variables and the manipulation test. 
 
\begin{figure}[ht]
    \centering
    \caption{Identification Assumption Check}
    \includegraphics[scale=0.4]{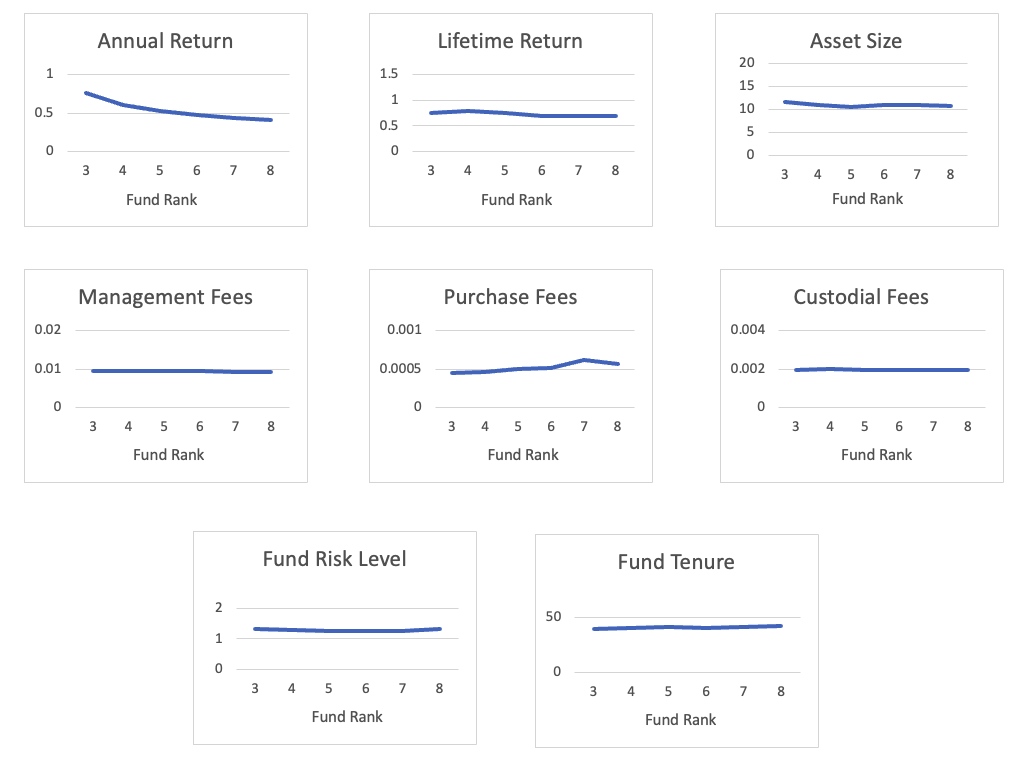}
    \label{fig:Identification}
\end{figure}

\textbf{Continuity of Observable Variables} One of the foundational assumptions underpinning RDD is that while factors like annual returns, management fees, purchase fees, or asset size, vary continuously around the cutoff, becoming a recommended fund to investors is discontinuous at the fifth rank in annual returns. We test this continuity assumption by plotting the average fund annual return, lifetime return, asset size, management fees, purchase fees, custodial fees, fund risk levels, and fund tenure by fund ranking from third to eighth during our sample period. Figure~\ref{fig:Identification} presents a visual representation of our results. We observe that annual returns decrease continuously and all the fund attributes also vary smoothly around the cutoff, thus supporting the continuity assumption.

\begin{figure}[ht]
    \centering
    \caption{Manipulation Test}
    \includegraphics[scale=0.4]{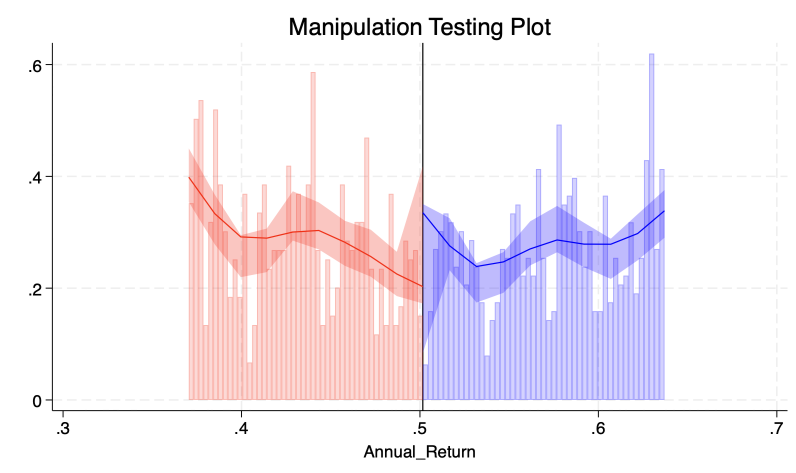}
    \label{fig:Manipulation}
\end{figure}

\textbf{Manipulation Test} Another challenge of RDD design is the manipulation of the running variable. If the running variable, i.e. annual return, is manipulated in our RDD design, the quasi-random assignment of a fund's recommendation status would be compromised and the running variable will change discontinuously at the cutoff, thus violating the continuity assumption and undermining the validity of our RDD-derived estimates. Given that annual returns, as opposed to other variables like management fees, are outside the direct control of the funds, they are less susceptible to manipulation. Consequently, it's reasonable to assume that funds cannot manipulate their recommendation status. To more rigorously test for manipulation, we perform the manipulation test on the annual return using the local polynomial density estimators proposed by \cite{cattaneo2020simple}. Results are presented in Figure~\ref {fig:Manipulation}, with the middle line being the cutoff point. Given the cutoff value that determines a fund's recommendation status changes daily, we use the average daily value of annual returns for funds ranked in fifth and sixth place to approximate the cutoff value. In Figure~\ref {fig:Manipulation}, we observe a large overlap of annual return density estimation in the middle. Results also show the annual return density difference around the cutoff to be insignificant ($p = 0.398$) and the change in density is smooth and continuous. This suggests there is no statistical evidence of systematic manipulation of annual returns.

\subsection{\label{purchase positive}Estimation Results}
Table~\ref{tab:mainEffect} contains the estimated effect of recommendations on online investors’ purchase behaviors. The estimates in Column (1) suggest that funds featured by product recommendations prompt more purchases and the effect is statistically significant (coefficient = 0.055; $p < 0.01$). The log-linear model indicates a 5.6\%\footnote{According to the log-linear model, the coefficient translates to 100*(exp(0.055) - 1)\% change.} increase in the number of purchases of recommended funds, compared with non-recommended funds. As shown in Columns (2) and (3), this result remains consistent when we gradually relax our model specification by removing control variables and fixed effects. This finding suggests that product recommendations that use historical returns exert influence in financial product domains. However, as prior literature has established, historical returns are not reliable quality signals in financial markets \citep{kahn1995does, jain2000truth, choi2020carhart}. As a result, investors still follow recommendations even though such recommendations based on historical return don't provide reliable quality signals from the perspective of rational investors.

\begin{table}[ht!]
\footnotesize
\caption{Effect of Recommendation on Fund Purchases}
  \centering
    \begin{tabular}{lcccc}
    \midrule
    \multicolumn{1}{l}{Dependent Variable:} & \multicolumn{1}{c}{(1)} & \multicolumn{1}{c}{(2)} & \multicolumn{1}{c}{(3)} \\
    \multicolumn{1}{l}{log(NumberOfPurchases)} & \multicolumn{1}{c}{} & \multicolumn{1}{c}{} & \multicolumn{1}{c}{} \\
    \midrule
    \multicolumn{1}{l}{Recommended} & \multicolumn{1}{c}{0.055***} & \multicolumn{1}{c}{0.058***} & \multicolumn{1}{c}{0.044***} \\
    \multicolumn{1}{r}{} & \multicolumn{1}{c}{(0.006)} & \multicolumn{1}{c}{(0.006)} & \multicolumn{1}{c}{(0.007)} \\
    \multicolumn{1}{l}{Polynomial Degree} & \multicolumn{1}{c}{2} & \multicolumn{1}{c}{2} & \multicolumn{1}{c}{2} \\
    \multicolumn{1}{l}{Fund Type Dummies} & \multicolumn{1}{c}{Yes} & \multicolumn{1}{c}{Yes} & \multicolumn{1}{c}{No} \\
     \multicolumn{1}{l}{Month Dummies} & \multicolumn{1}{c}{Yes} & \multicolumn{1}{c}{Yes} & \multicolumn{1}{c}{No} \\
      \multicolumn{1}{l}{Day of Week Dummies} & \multicolumn{1}{c}{Yes} & \multicolumn{1}{c}{Yes} & \multicolumn{1}{c}{No} \\
    \multicolumn{1}{l}{Control Variables} & \multicolumn{1}{c}{Yes} & \multicolumn{1}{c}{No} & \multicolumn{1}{c}{No} \\
    \multicolumn{1}{l}{Number of Observations} & \multicolumn{1}{c}{21,834} & \multicolumn{1}{c}{21,834} & \multicolumn{1}{c}{21,834} \\
    \multicolumn{1}{l}{\textcolor[rgb]{ .133,  .133,  .133}{$R^2$}} & \multicolumn{1}{c}{0.153}  & \multicolumn{1}{c}{0.081} & \multicolumn{1}{c}{0.010} \\
    \midrule
    \multicolumn{4}{p{28em}}{\scriptsize Notes: {***} $p<0.01$, {**} $p<0.05$, {*} $p<0.1$. The dependent variable in Columns (1) - (3) is log(NumberOfPurchases). Control variables include management fees, purchase fees, custodial fees, asset size, dummy for valuable fund brand, fund past returns, fund risk level, and fund tenure. Robust standard errors are in parentheses.} \\
    \end{tabular}
  \label{tab:mainEffect}
\end{table}

\subsection{\label{robustness}Robustness Checks}
To strengthen the causal interpretation of our results, we conduct several robustness checks, including (1) a falsification exercise, (2) alternative bandwidths, (3) alternative measurements, and (4) a different model specification. Robustness check results are presented in Appendix~\ref{B} and they align with our main findings, as we detail next. 

\subsubsection{\label{falsification}Falsification Exercise}
To rule out an alternative explanation that any funds with higher rankings prompt more purchases, we perform a falsification exercise in which funds that rank fourth are the recommended funds and funds that rank fifth represent the ``placebo'' non-recommended funds. Since both funds are recommended, funds that rank fourth should not account for significantly more purchases than funds that rank fifth. The results, as presented in Column (1) of Table~\ref{tab:falsificationAlternateBandwidth} in Appendix~\ref{B}, indicate that this placebo recommendation effect is not statistically significant. We perform a similar placebo assessment of funds that rank in sixth and seventh place. As the results in Column (2) of Table~\ref{tab:falsificationAlternateBandwidth} in Appendix~\ref{B} show, the effect of ``placebo'' product recommendations remains insignificant. The falsification test thus strengthens our causal interpretation of the main effect estimation.

\subsubsection{\label{bandwidth}Alternative Bandwidths}
Because funds that are closer in rank might be more similar than funds that are farther in rank, we check the robustness of our results with a narrower bandwidth. That is, we include only the funds that rank in fifth place as recommended and those that rank in sixth place as non-recommended. The estimation result, provided in Column (3) of Table~\ref{tab:falsificationAlternateBandwidth} in Appendix~\ref{B}, shows that the recommendation effect on the number of purchases remains positive and statistically significant. We also perform similar analyses by including funds that rank in fourth and fifth place as recommended funds and funds that rank in sixth and seventh place as non-recommended funds. Results are provided in Column (4) of Table~\ref{tab:falsificationAlternateBandwidth} in Appendix~\ref{B} and the results are positive and statistically significant. Thus, our results are robust to the choice of narrower bandwidths.

\subsubsection{\label{moredv}Alternative Measurements}
Besides the number of purchases, there are alternative measures to represent the sales of funds such as the number of unique buyers, the total units of the fund sold, and the cash amount of the fund sold. To test the sensitivity of our dependent variable, we use these three alternative measurements to examine the effect of product recommendations on fund sales.

Specifically, $\log{(NumberOfInvestors_{it})}$ is the logarithm of the number of unique investors who purchase fund $i$ on day $t$; $\log(PurchaseAmount_{it})$ is the logarithm of the total units of fund $i$ purchased on day $t$; $MoneyAmount_{it}$ is the inflow of funds in cash amount for fund $i$ on day $t$. Our regression model is the same as in Equation~\ref{eq:mainEquation} but we replace the dependent variable with these three alternative measurements. The estimation results are presented in Table~\ref{tab:altDV} in Appendix~\ref{B}. The coefficients of $Recommended_{it}$ are all positive and significant in these three models, in line with our primary results.

\subsubsection{\label{poisson}Different Model Specification}
Since the dependent variable log($NumberOfPurchases_{it}$) in our study is count data, to test the sensitivity of model specification, we use the Poisson regression model instead of the linear model to check whether our results remain consistent \citep{consul1992generalized}. We present the results in Column (1) of Table~\ref{tab:altModelSubsample} in Appendix~\ref{B} where we use the same formula as in Equation~\ref{eq:mainEquation}. The coefficient for $Recommended_{it}$ remains positive and statistically significant.

\section{\label{investor level effect}Effect of Product Recommendations by Investor Socioeconomic Status}
The results thus far show that on average, product recommendations affect investors' purchase behaviors. However, different investors might perceive quality signals provided by product recommendations differently. Therefore, in this section, we examine the heterogeneous effects of product recommendations across investors with different socioeconomic statuses (SES). 

\subsection{\label{segmentation}Investor Segmentation}
In our study, we collect investors' information on income, education, and liability,\footnote{The summary statistics of these variables are presented in Table~\ref{tab:investorSummary} in Appendix A} which are the three most essential characteristics to determine investors' SES level \citep{calvet2009measuring}. In detail, each investor has a score for income, education, and liability respectively. These scores range from 0 to 2, with a higher value indicating a higher level for variables of income and education. To make it consistent (i.e. a higher value means better wellness), a higher value indicates a lower level for the variable of liability. We sum up the scores of these three variables for each investor to represent the overall SES score. Then we divide investors into low, medium, and high SES levels by using cutoffs at 1/3 and 2/3 of the highest sum of the scores. A summary of the three investor groups, characterized by varying levels of SES, appears in Table~\ref{tab:investorSegmentation}.

\begin{table}[ht!]
\footnotesize
\caption{Investor Segmentation}
  \centering
    \begin{tabular}{ccccc}
    \midrule
    Investors Group & \multicolumn{1}{c}{\# of Investors} & \multicolumn{1}{c}{Income} & \multicolumn{1}{c}{Education} & \multicolumn{1}{c}{Liability} \\
    \midrule
    \makecell{Low SES}  & 9,657 & 0.035 & 0.022 & 0.978 \\
    \midrule
    \makecell{Medium SES} & 1,855 & 1.075 & 1.009 & 1.380 \\
    \midrule
   \makecell{High SES} & 929 & 1.904 & 1.273 & 2.000 \\
    \midrule
    \multicolumn{5}{p{30em}}{\scriptsize Notes: Investor segmentation is based on the combined scores of investors' income, education, and liability levels. Specific values for the scores are presented in Table~\ref{tab:investorSummary}. Investors are segmented into three groups using cutoffs at 1/3 and 2/3 of the highest sum of the scores. Columns (3) - (5) presents the average income, education, and liability scores by investor SES group.} \\
    \end{tabular}
  \label{tab:investorSegmentation}
\end{table}

\subsection{\label{investor heterogeneous effect}Heterogeneous Effects of Product Recommendations by Investor Socioeconomic Status}
Since investors with diverse SES might perceive the quality signal from product recommendations differently, here we examine how the effect of product recommendations varies across investor SES. We perform the analyses similar to Equation~\ref{eq:mainEquation} but replace our dependent variable with log($NumberOfPurchases_{ijt}$), the logarithm of the total number of purchases for fund $i$ on the day $t$ from investor group $j$. Group $j$ refers to investors of low, medium, and high SES respectively. The estimation results are presented in Table~\ref{tab:recommendByGroup}. Columns (1) - (3) show the estimation results by using sub-samples of investors with low, medium, and high SES respectively. Results suggest that the purchase behaviors of all three investor groups are significantly influenced by recommendations provided by the platform despite varying degrees of magnitude. The recommendation effect is strongest for investors with low SES, as indicated in Column (1), suggesting that they are more likely to purchase recommended funds. On average, low SES investors engage in 4.1\% more recommended fund purchases than non-recommended funds. With the Chow test, we find that this number is significantly larger than investors with medium-level ($F = 5.04; p < 0.05$) and high-level SES ($F = 8.68; p < 0.01$). Low SES investors are least likely to possess enough knowledge to choose funds and make rational investment decisions. As a result, compared with medium and high SES investors, investors of low SES are more likely to perceive recommendations by the platform as a strong quality signal that should inform their purchase decision.

\begin{table}[ht!]
\footnotesize
\caption{Heterogeneous Effects of Recommendation by Investor SES}
  \centering
    \begin{tabular}{lccc}
    \midrule
    \multicolumn{1}{l}{Dependent Variable:} & \multicolumn{1}{c}{(1)} & \multicolumn{1}{c}{(2)} & \multicolumn{1}{c}{(3)} \\
    \multicolumn{1}{l}{log(NumberOfPurchases)} & \multicolumn{1}{c}{Low SES} & \multicolumn{1}{c}{Med SES} & \multicolumn{1}{c}{High SES}\\
    \midrule
    \multicolumn{1}{l}{Recommended} & \multicolumn{1}{c}{0.040***} & \multicolumn{1}{c}{0.020***} & \multicolumn{1}{c}{0.016***} \\
     \multicolumn{1}{r}{} & \multicolumn{1}{c}{(0.005)} & \multicolumn{1}{c}{(0.003)} & \multicolumn{1}{c}{(0.003)} \\
    \multicolumn{1}{l}{Polynomial Degree} & \multicolumn{1}{c}{2} & \multicolumn{1}{c}{2} & \multicolumn{1}{c}{2} \\
    \multicolumn{1}{l}{Fund Type Dummies} & \multicolumn{1}{c}{Yes} & \multicolumn{1}{c}{Yes} & \multicolumn{1}{c}{Yes} \\
    \multicolumn{1}{l}{Month Dummies} & \multicolumn{1}{c}{Yes} & \multicolumn{1}{c}{Yes} & \multicolumn{1}{c}{Yes} \\
    \multicolumn{1}{l}{Day of Week Dummies} & \multicolumn{1}{c}{Yes} & \multicolumn{1}{c}{Yes} & \multicolumn{1}{c}{Yes} \\
    \multicolumn{1}{l}{Control Variables} & \multicolumn{1}{c}{Yes} & \multicolumn{1}{c}{Yes} & \multicolumn{1}{c}{Yes} \\
    \multicolumn{1}{l}{Number of Observations} & \multicolumn{1}{c}{21,834} & \multicolumn{1}{c}{21,834} & \multicolumn{1}{c}{21,834} \\
    \multicolumn{1}{l}{\textcolor[rgb]{ .133,  .133,  .133}{$R^2$}} & \multicolumn{1}{c}{0.147} & \multicolumn{1}{c}{0.079} & \multicolumn{1}{c}{0.070} \\
    \midrule
    \multicolumn{4}{p{28em}}{\scriptsize Notes: {***} $p<0.01$, {**} $p<0.05$, {*} $p<0.1$. The dependent variable in Columns (1) - (3) is log(NumberOfPurchases). Control variables include management fees, purchase fees, custodial fees, asset size, dummy for valuable fund brand, fund past returns, fund risk level, and fund tenure. Columns (1) - (3) include analysis results using sub-samples of investors with low, medium, and high SES respectively. Robust standard errors are in parentheses.} \\
    \end{tabular}
  \label{tab:recommendByGroup}
\end{table}

\section{\label{effect on return}Effect of Following Product Recommendations on Investment Return}
Our study indicates that investors follow product recommendations using historical returns. However, existing literature establishes that historical returns cannot reliably predict future returns \citep{kahn1995does, clifford2021salience, barber2021attention}. Therefore, we question whether purchasing recommended funds leads to better or worse investment return outcomes for investors. In this section, we examine the effect of following product recommendations on investment returns.

\subsection{\label{negative effect}Investment Return Outcome after Purchasing Recommended versus Non-Recommended Funds}
To compare the investment return outcomes after investors purchase recommended versus non-recommended funds, we construct our dependent variable $InvestReturn_{ijt}$ as follows. Since an investment consists of multiple purchases and redemption, we cannot measure return using a single transaction and we instead define an investment session to calculate the investment return. Each investment session starts with an investor purchasing a fund. After that purchase, we monitor the investors' follow-up purchases, redemption, and the entire net holding history of the purchased fund in the investment session. The investment session concludes when the net holdings of the purchased fund hit zero, with the final investment return calculated using all the transactions in the session.\footnote{For investment sessions that do not end on the last day of the sample period, we presume all remaining funds are sold on December 31, 2016, our sample's final date. That gives us an investor's paper loss for that investment session. In the following models, we also include an indicator to show whether the investment session ends before the last day of the sample period.} In summary, $InvestReturn_{ijt}$ is calculated as
\begin{eqnarray} \label{eq:invest_return_var}
\begin{aligned}
InvestReturn_{ijt} = \frac{\sum_T NAV_T * Amt_T * Redemption_T - \sum_T NAV_T * Amt_T * Purchase_T  }{\sum_T NAV_T * Amount_T * Purhcase_T}
\end{aligned}
\end{eqnarray}

The subscript T covers all transactions within the investment session. $NAV_T$ is the net asset value of the fund and $Amt_T$ is the units of funds involved in the transaction. $Redemption_T$ and $Purchase_T$ are dummies denoting whether the transaction is a fund purchase or redemption. The final investment returns are presented in basis points. Our unit of analysis is fund ($i$) - investor ($j$) - day ($t$). $InvestReturn_{ijt}$ denotes the investment return for an investment session starting with fund $i$ being purchased by investor $j$ on day $t$.  If the fund is recommended on the day it is first purchased, then we define the fund in this investment session as a recommended fund. 

To investigate the investment return difference between recommended and non-recommended funds, we adopt a similar RDD approach in Section \ref{main effect} by comparing the funds ranked third to fifth with those ranked sixth to eighth. We use the following model to formally estimate the effect of following product recommendations on the investment return:
\begin{eqnarray} \label{eq:invest_return}
\begin{aligned}
InvestReturn_{ijt} = &\beta\ast Recommended_{it}+Group_j+ f(AnnualReturn_{it})+ \\
& \theta\ast X_{it}+\alpha_{i}+\gamma_{t}+\xi_{t}+\epsilon_{ijt}
\end{aligned}
\end{eqnarray}
$Group_j$ are the investor SES group dummies. The fixed effects and control variables are the same as in Equation~\ref{eq:mainEquation}.

The estimation results, presented in Column (1) of Table~\ref{tab:recommendInvestReturn}, suggest that purchasing recommended funds is associated with 0.89 lower basis points in returns. Such a difference is also statistically significant ($p < 0.05$). On average, investors appear to perform worse when they purchase recommended funds (versus non-recommended funds). This finding is in direct conflict with the investment platforms' goals of using recommendations to help investors discover good financial products and achieve higher investment returns.\footnote{\url{https://www.investopedia.com/best-online-brokers-4587872}} Instead of helping investors, investors actually perform worse when following product recommendations.

\begin{table}[ht!]
\footnotesize
  \centering
  \caption{Effect of Recommendation on Investment Return}
    \begin{tabular}{lcccc}
    \midrule
    \multicolumn{1}{l}{Dependent Variable:} & \multicolumn{1}{c}{(1)} & \multicolumn{1}{c}{(2)} & \multicolumn{1}{c}{(3)} & \multicolumn{1}{c}{(4)} \\
    \multicolumn{1}{l}{InvestReturn} & \multicolumn{1}{c}{All Investors} & \multicolumn{1}{c}{Low SES} & \multicolumn{1}{c}{Med SES} & \multicolumn{1}{c}{High SES}\\
    \midrule
    \multicolumn{1}{l}{Recommended} & \multicolumn{1}{c}{-0.895**} & \multicolumn{1}{c}{-1.127**} & \multicolumn{1}{c}{-1.105} & \multicolumn{1}{c}{1.597} \\
    \multicolumn{1}{r}{} & \multicolumn{1}{c}{(0.452)} & \multicolumn{1}{c}{(0.463)} & \multicolumn{1}{c}{(0.793)} & \multicolumn{1}{c}{(1.979)} \\
    \multicolumn{1}{l}{Polynomial Degree} & \multicolumn{1}{c}{2} & \multicolumn{1}{c}{2} & \multicolumn{1}{c}{2} & \multicolumn{1}{c}{2} \\
    \multicolumn{1}{l}{Investor Group Dummies} & \multicolumn{1}{c}{Yes} & \multicolumn{1}{c}{No} & \multicolumn{1}{c}{No} & \multicolumn{1}{c}{No} \\
    \multicolumn{1}{l}{Fund Type Dummies} & \multicolumn{1}{c}{Yes} & \multicolumn{1}{c}{Yes} & \multicolumn{1}{c}{Yes} & \multicolumn{1}{c}{Yes} \\
    \multicolumn{1}{l}{Month Dummies} & \multicolumn{1}{c}{Yes} & \multicolumn{1}{c}{Yes} & \multicolumn{1}{c}{Yes} & \multicolumn{1}{c}{Yes} \\
    \multicolumn{1}{l}{Day of Week Dummies} & \multicolumn{1}{c}{Yes} & \multicolumn{1}{c}{Yes} & \multicolumn{1}{c}{Yes} & \multicolumn{1}{c}{Yes} \\
    \multicolumn{1}{l}{Control Variables} & \multicolumn{1}{c}{Yes} & \multicolumn{1}{c}{Yes} & \multicolumn{1}{c}{Yes} & \multicolumn{1}{c}{Yes} \\
    \multicolumn{1}{l}{Number of Observations} & \multicolumn{1}{c}{4,618} & \multicolumn{1}{c}{3,221} & \multicolumn{1}{c}{886} & \multicolumn{1}{c}{511} \\
    \multicolumn{1}{l}{\textcolor[rgb]{ .133,  .133,  .133}{\textit{$R^2$}}} & \multicolumn{1}{c}{0.135} & \multicolumn{1}{c}{0.118} & \multicolumn{1}{c}{0.193} & \multicolumn{1}{c}{0.491}\\
    \midrule
    \multicolumn{5}{p{36em}}{\scriptsize Notes: {***} $p<0.01$, {**} $p<0.05$, {*} $p<0.1$. The dependent variable in Columns (1) - (4) is InvestReturn. Control variables include management fees, purchase fees, custodial fees, asset size, dummy for valuable fund brand, fund past returns, fund risk level, and fund tenure. Columns (2)-(4) show the estimation results for the sub-samples of investors with low, medium, and high SES respectively. Robust standard errors are in parentheses.} \\
    \end{tabular}%
  \label{tab:recommendInvestReturn}
\end{table}%

\subsection{\label{investor performance}Heterogeneous Effect of Product Recommendations on Investment Return by Investor SES}
Our previous results consistently indicate that the product recommendations effect is heterogeneous across investors with different SES. These heterogeneous effects in turn might lead to heterogeneous investment returns across investors with different SES. We examine such heterogeneous effects using Equation~\ref{eq:invest_return} and perform sub-sample analysis, only considering investment sessions conducted by investors with low, medium, and high SES respectively. 

The estimation results are presented in Columns (2) - (4) of Table~\ref{tab:recommendInvestReturn}. The results show that only investors with low SES suffer significantly worse performance when purchasing recommended funds. For investors of medium and high SES, purchasing recommended funds is not associated with significantly worse returns than purchasing non-recommended funds. Along with our prior results that investors with low SES are more likely to purchase recommended funds, these findings imply that low SES investors rely most on product recommendations and are also hurt the most by recommendations. This potential dark side of product recommendations is worrisome, in that investors who need recommendations the most also suffer the most, which could amplify the problematic wealth inequality among investors in financial markets.

\section{\label{return explain}Subsequent Behavioral Changes and Potential Explanations on Worse Performance When Following Recommendations}
In this section, we delve into investors' subsequent behavioral changes after following product recommendations and provide three potential explanations, supported by empirical data, as to why purchasing recommended funds often results in sub-optimal performance, especially for low SES investors. First, we examine search behavior differences in the purchase stage, comparing search efforts when investors purchase recommended versus non-recommended funds. Then we investigate the difference in fund redemption timing between investors who purchase recommended funds and those who purchase non-recommended funds. Finally, we evaluate and compare the future performance between recommended and non-recommended funds as another possibility.

\subsection{\label{fund research}Behavioral Differences in the Purchase Stage}
When investors apply greater effort, they are more likely to make informed and rational investment choices \citep{barberis2003survey}. However, the presence of product recommendations might lead investors to diminish their search efforts, potentially culminating in less optimal purchasing decisions. Thus, it is critical to compare differences in search behaviors when investors purchase recommended and non-recommended funds. Using investors' individual-level click stream data, we examine the impact of following product recommendations on fund search efforts.

\subsubsection{\label{fund search effort all}Overall Search Efforts in the Purchase Stage}
We use the total search time and size of the fund consideration set to represent investors' search efforts respectively. To calculate the total search time, we monitor the fund click history of investors and use the temporal gap between successive clicks to estimate the time devoted to each individual fund measured in seconds. Then we calculate the total search time as the cumulative duration an investor devotes to all funds within the three days preceding any fund purchases. To examine the impact of following fund recommendations on investors' total search time, we adopt the model similar to Equation ~\ref{eq:invest_return} but substitute the dependent variable with the total search time. The results presented in Column (1) of Table~\ref{tab:recommendSearchEffort} reveal that investors tend to spend significantly less total search time when purchasing recommended funds as opposed to non-recommended ones. This suggests that product recommendations may inadvertently curtail the thoroughness of investors' overall search efforts. 

Next, we examine how investors' fund consideration set size changes when following fund recommendations. Specifically, we measure consideration set size by counting the number of unique funds investors engage with three days prior to making a purchase. We adopt the model similar to Equation~\ref{eq:invest_return} and the findings in Column (2) of Table~\ref{tab:recommendSearchEffort} indicate that, on average, there's no statistically significant variation in the number of distinct funds investors consider when purchasing recommended funds compared with non-recommended ones. This implies that while product recommendations tend to diminish the time investors spend on research, they don't notably influence the size of funds within an investor's consideration set.

As investors decrease their total search time but do not change the consideration set size when following product recommendations, it is expected that the average time investors dedicate to each individual fund also decreases. To formally test this conjecture, we construct the average search time per fund by dividing the total search time by the fund consideration set size. We adopt the model similar to Equation~\ref{eq:invest_return} and our findings in Column (3) of Table~\ref{tab:recommendSearchEffort} reveal that, when following fund recommendations, investors spend around one minute less average search time on each fund.

\begin{table}[ht!]
\footnotesize
  \centering
  \caption{Effect of Recommendation on Fund Search Effort}
    \begin{tabular}{lccc}
    \midrule
    \multicolumn{1}{l}{Dependent Variable:} & \multicolumn{1}{c}{(1)} & \multicolumn{1}{c}{(2)} & \multicolumn{1}{c}{(3)}\\
    \multicolumn{1}{l}{} & \multicolumn{1}{c}{Total Time} & \multicolumn{1}{c}{Unique \# of Funds} & \multicolumn{1}{c}{Avg Time Per Fund} \\
    \midrule
    \multicolumn{1}{l}{Recommended} & \multicolumn{1}{c}{-1332.860*} & \multicolumn{1}{c}{-0.048} & \multicolumn{1}{c}{-65.471**} \\
    \multicolumn{1}{r}{} & \multicolumn{1}{c}{(755.645)} & \multicolumn{1}{c}{(0.376)} & \multicolumn{1}{c}{(30.841)} \\
    \multicolumn{1}{l}{Polynomial Degree} & \multicolumn{1}{c}{2} & \multicolumn{1}{c}{2} & \multicolumn{1}{c}{2} \\
    \multicolumn{1}{l}{Investor Group Dummies} & \multicolumn{1}{c}{Yes} & \multicolumn{1}{c}{Yes} & \multicolumn{1}{c}{Yes} \\
    \multicolumn{1}{l}{Fund Type Dummies} & \multicolumn{1}{c}{Yes} & \multicolumn{1}{c}{Yes} & \multicolumn{1}{c}{Yes} \\
    \multicolumn{1}{l}{Month Dummies} & \multicolumn{1}{c}{Yes} & \multicolumn{1}{c}{Yes} & \multicolumn{1}{c}{Yes} \\
    \multicolumn{1}{l}{Day of Week Dummies} & \multicolumn{1}{c}{Yes} & \multicolumn{1}{c}{Yes} & \multicolumn{1}{c}{Yes} \\
    \multicolumn{1}{l}{Control Variables} & \multicolumn{1}{c}{Yes} & \multicolumn{1}{c}{Yes} & \multicolumn{1}{c}{Yes} \\
    \multicolumn{1}{l}{Number of Observations} & \multicolumn{1}{c}{5,339} & \multicolumn{1}{c}{5,339} & \multicolumn{1}{c}{5,339} \\
    \multicolumn{1}{l}{\textcolor[rgb]{ .133,  .133,  .133}{\textit{$R^2$}}} & \multicolumn{1}{c}{0.055} & \multicolumn{1}{c}{0.050} & \multicolumn{1}{c}{0.067} \\
    \midrule
    \multicolumn{4}{p{36em}}{\scriptsize Notes: {***} $p<0.01$, {**} $p<0.05$, {*} $p<0.1$. The dependent variables in Columns (1) - (3) are the search time spent on all funds, the average search time per fund, and the unique number of funds viewed. Control variables include management fees, purchase fees, custodial fees, asset size, dummy for valuable fund brand, fund past returns, fund risk level, and fund tenure. Robust standard errors are in parentheses.} \\
    \end{tabular}%
  \label{tab:recommendSearchEffort}
\end{table}%
\subsubsection{\label{fund search effort allocate}Variations of Search Efforts Reduction by SES}
Our earlier results indicate that investors of different SES interpret product recommendation quality signals differently. Consequently, the effects of product recommendations on fund search efforts could also differ across these groups. We explore this differential impact using sub-sample analysis by investor SES groups and use the same approach in Section ~\ref{fund search effort all}, with the dependent variable being the average search time per fund. Results are summarized In Table ~\ref{tab:recommendSearchEffortSES}. Column (1) indicates that investors of low SES spend significantly less time for each fund when following product recommendations, with the coefficient $Recommended_{it}$ being negative and statistically significant. However, as shown in Columns (2) and (3), the effect is insignificant for investors with medium and high SES, suggesting that only low SES investors' search efforts are negatively impacted by product recommendations. This could potentially explain our results in Section~\ref{investor performance} which finds that only investors of low SES realize significantly worse investment returns when following product recommendations.

Employing the Chow test, we further discern that the influence of recommendations is significantly more pronounced for low SES investors compared to both medium SES ($F = 10.30; p < 0.01$) and high SES investors ($F = 5.21; p < 0.05$). This distinction reveals that low SES investors exhibit a considerably steeper decline in search effort when choosing recommended over non-recommended funds.

This observation is in line with \cite{havakhor2021tech}, which posits that increased access to tech-enabled financial performance data can induce gambling-like behaviors among less sophisticated investors. Such accessibility to vast performance data might amplify overconfidence and trading frequency among these investors, primarily because they often struggle to effectively process relevant financial details. Our study supports this conclusion with novel empirical evidence that when presented with recommendations spotlighting historically high-performing funds, less adept investors might be unduly swayed, manifesting increased overconfidence and impulsivity in trading. This behavioral pattern resonates with our earlier conclusions that low SES investors are more likely to follow product recommendations and realize significantly worse investment returns.

\begin{table}[ht!]
\footnotesize
  \centering
  \caption{Effect of Recommendation on Fund Search Effort by Investor SES}
    \begin{tabular}{lccc}
    \midrule
    \multicolumn{1}{l}{Dependent Variable:} & \multicolumn{1}{c}{(1)} & \multicolumn{1}{c}{(2)} & \multicolumn{1}{c}{(3)}\\
    \multicolumn{1}{l}{Avg Time Per Fund} & \multicolumn{1}{c}{Low SES} & \multicolumn{1}{c}{Med SES} & \multicolumn{1}{c}{High SES} \\
    \midrule
    \multicolumn{1}{l}{Recommended} & \multicolumn{1}{c}{-81.831**} & \multicolumn{1}{c}{14.068} & \multicolumn{1}{c}{11.396} \\
    \multicolumn{1}{r}{} & \multicolumn{1}{c}{(38.918)} & \multicolumn{1}{c}{(53.648)} & \multicolumn{1}{c}{(61.256)} \\
    \multicolumn{1}{l}{Polynomial Degree} & \multicolumn{1}{c}{2} & \multicolumn{1}{c}{2} & \multicolumn{1}{c}{2} \\
    \multicolumn{1}{l}{Investor Group Dummies} & \multicolumn{1}{c}{Yes} & \multicolumn{1}{c}{Yes} & \multicolumn{1}{c}{Yes} \\
    \multicolumn{1}{l}{Fund Type Dummies} & \multicolumn{1}{c}{Yes} & \multicolumn{1}{c}{Yes} & \multicolumn{1}{c}{Yes} \\
    \multicolumn{1}{l}{Month Dummies} & \multicolumn{1}{c}{Yes} & \multicolumn{1}{c}{Yes} & \multicolumn{1}{c}{Yes} \\
    \multicolumn{1}{l}{Day of Week Dummies} & \multicolumn{1}{c}{Yes} & \multicolumn{1}{c}{Yes} & \multicolumn{1}{c}{Yes} \\
    \multicolumn{1}{l}{Control Variables} & \multicolumn{1}{c}{Yes} & \multicolumn{1}{c}{Yes} & \multicolumn{1}{c}{Yes} \\
    \multicolumn{1}{l}{Number of Observations} & \multicolumn{1}{c}{4,143} & \multicolumn{1}{c}{609} & \multicolumn{1}{c}{587} \\
    \multicolumn{1}{l}{\textcolor[rgb]{ .133,  .133,  .133}{\textit{$R^2$}}} & \multicolumn{1}{c}{0.074} & \multicolumn{1}{c}{0.197} & \multicolumn{1}{c}{0.088} \\
    \midrule
    \multicolumn{4}{p{28em}}{\scriptsize Notes: {***} $p<0.01$, {**} $p<0.05$, {*} $p<0.1$. The dependent variable in Columns (1) - (3) is the average search time per fund. Columns (1) - (3) include analysis results using sub-samples of investors with low, medium, and high SES respectively. Control variables include management fees, purchase fees, custodial fees, asset size, dummy for valuable fund brand, fund past returns, fund risk level, and fund tenure. Robust standard errors are in parentheses.} \\
    \end{tabular}%
  \label{tab:recommendSearchEffortSES}
\end{table}%

\subsubsection{\label{fund search effort specific}Search Efforts Allocation in the Purchase Stage}
Building on our analysis that overall search efforts decrease preceding the purchase of recommended funds compared with non-recommended ones, we want to understand what kind of funds get less search time by investors. To address this, we use the model analogous to Table~\ref{tab:recommendSearchEffort} with the dependent variables replaced by the average search time allocated to funds that investors eventually purchase and to the funds that they do not purchase. Results are presented in Table~\ref{tab:recommendSearchCategory}.  In Column (1), our findings indicate preceding the fund purchase, there is no significant difference between the average search time allocated to the purchased recommended and the purchased non-recommended fund. Transitioning our focus to the average time dedicated to funds investors eventually bypass, results from Column (2) reveal that when investors follow fund recommendations, they markedly reduce the time spent considering alternative fund options. This suggests that recommendations play a significant role in diminishing investors' exploration of non-purchased funds.

As the search efforts in alternative funds reduce when following product recommendations, we further probe into the average time investors allocate to non-purchased funds. Specifically, we examine the time devoted to non-purchased funds within the same fund type and different fund types as the purchased fund respectively. We use the model similar to Table~\ref{tab:recommendSearchEffort}, and interestingly, our results in Column (3) reveal that when following fund recommendations, investors significantly curtail their research time for other funds within the same fund type. This finding is consistent with the design of product recommendations that highlight top-performing funds within each fund type. Hence, when a recommended fund is purchased, investors tend to allocate less time to investigating its similar counterparts. In a contrasting analysis, detailed in Column (4), we don't find a significant change in search time dedicated to funds in different fund types when purchasing recommended (versus non-recommended funds).

\begin{table}[ht!]
\footnotesize
  \centering
  \caption{Effect of Recommendation on Fund Search Time by Fund Category}
    \begin{tabular}{lcccc}
    \midrule
    \multicolumn{1}{l}{Dependent Variable:} & \multicolumn{1}{c}{(1)} & \multicolumn{1}{c}{(2)} & \multicolumn{1}{c}{(3)} & \multicolumn{1}{c}{(4)} \\
    \multicolumn{1}{l}{} & \multicolumn{1}{c}{\makecell{Purchased \\ Fund}} & \multicolumn{1}{c}{\makecell{Non-Purchased \\ Funds}} & \multicolumn{1}{c}{\makecell{Same Type \\ Funds}} & \multicolumn{1}{c}{\makecell{Different Type \\ Funds}} \\
    \midrule
    \multicolumn{1}{l}{Recommended} & \multicolumn{1}{c}{-115.246} & \multicolumn{1}{c}{-48.259*} & \multicolumn{1}{c}{-45.434**} & \multicolumn{1}{c}{-68.528} \\
    \multicolumn{1}{r}{} & \multicolumn{1}{c}{(70.726)} & \multicolumn{1}{c}{(28.197)} & \multicolumn{1}{c}{(21.924)} & \multicolumn{1}{c}{(44.735)} \\
    \multicolumn{1}{l}{Polynomial Degree} & \multicolumn{1}{c}{2} & \multicolumn{1}{c}{2} & \multicolumn{1}{c}{2}  & \multicolumn{1}{c}{2} \\
    \multicolumn{1}{l}{Investor Group Dummies} & \multicolumn{1}{c}{Yes} & \multicolumn{1}{c}{Yes} & \multicolumn{1}{c}{Yes} & \multicolumn{1}{c}{Yes} \\
    \multicolumn{1}{l}{Fund Type Dummies} & \multicolumn{1}{c}{Yes} & \multicolumn{1}{c}{Yes} & \multicolumn{1}{c}{Yes} & \multicolumn{1}{c}{Yes} \\
    \multicolumn{1}{l}{Month Dummies} & \multicolumn{1}{c}{Yes} & \multicolumn{1}{c}{Yes} & \multicolumn{1}{c}{Yes} & \multicolumn{1}{c}{Yes} \\
    \multicolumn{1}{l}{Day of Week Dummies} & \multicolumn{1}{c}{Yes} & \multicolumn{1}{c}{Yes} & \multicolumn{1}{c}{Yes} & \multicolumn{1}{c}{Yes} \\
    \multicolumn{1}{l}{Control Variables} & \multicolumn{1}{c}{Yes} & \multicolumn{1}{c}{Yes} & \multicolumn{1}{c}{Yes} & \multicolumn{1}{c}{Yes} \\
    \multicolumn{1}{l}{Number of Observations} & \multicolumn{1}{c}{5,339} & \multicolumn{1}{c}{5,339} & \multicolumn{1}{c}{5,339} & \multicolumn{1}{c}{5,339} \\
    \multicolumn{1}{l}{\textcolor[rgb]{ .133,  .133,  .133}{\textit{$R^2$}}} & \multicolumn{1}{c}{0.042} & \multicolumn{1}{c}{0.064} & \multicolumn{1}{c}{0.042} & \multicolumn{1}{c}{0.059} \\
    \midrule
    \multicolumn{5}{p{40em}}{\scriptsize Notes: {***} $p<0.01$, {**} $p<0.05$, {*} $p<0.1$. The dependent variables in Columns (1) - (4) are the average search time spent on the purchased fund, non-purchased funds, funds of the same type as the purchased fund, and funds of different types as the purchased fund. Control variables include management fees, purchase fees, custodial fees, asset size, the dummy for valuable fund brand, fund past returns, fund risk level, and fund tenure. Robust standard errors are in parentheses.} \\
    \end{tabular}%
  \label{tab:recommendSearchCategory}
\end{table}%

\subsection{\label{fund redemption timing}Behavioral Differences in Fund Redemption Timing}
The timing of fund redemption significantly influences investment returns. Typically, selling funds during price peaks yields superior returns. We investigate the possibility that investors who choose recommended funds, potentially due to lack of research as we show in Section~\ref{fund research}, possess weakened confidence in their decisions. Such weakened confidence might enable them to cash out their investments at sub-optimal times, thus worsening their returns.

To evaluate this possibility empirically, we undertake a simulation analyzing the effects of varying fund redemption timings on investment returns. In particular, we simulate scenarios where investors redeem their funds three months after their actual final redemption date in the investment session. Subsequently, we compute the disparity in fund return when redeeming on the simulated and actual final redemption date, which we term ``hypothetical excess return.'' A larger value in this metric signifies a worse redemption timing.

We perform a comparative analysis of the average hypothetical excess return between purchases of recommended and non-recommended funds. Results are presented in Table~\ref{tab:fundReturnCompareGroup}. Row (1) includes the average simulated hypothetical excess return for all investors in our sample and results indicate that, on average, those who purchase recommended funds would achieve higher hypothetical excess returns than those opting for non-recommended ones. This implies that investors who choose recommended funds tend to redeem at a worse time than those who purchase non-recommended funds. 

Since the effects of product recommendations are shown to be affected by investor SES levels, here we aim to further compare hypothetical excess return differences by investor SES levels. We present results at Rows (2) - (4) in Table~\ref{tab:fundReturnCompareGroup}. Intriguingly, our analysis reveals that the hypothetical excess returns for purchasing recommended funds are significantly positive only among investors from low SES backgrounds (0.036 vs 0.010; difference = $0.026; p < 0.01$), indicating that only low SES investors tend to redeem at a worse time when purchasing recommended funds (versus non-recommended funds). This marked discrepancy suggests that sub-optimal timing during the redemption phase could be a consequential factor that exacerbates poorer investment returns specifically for low SES investors who purchase recommended funds. This segment of investors is disproportionately impacted by less-than-ideal redemption timing and corresponds with our prior results that only investors of low SES realize significantly worse investment returns when following fund recommendations. The finding contributes to our understanding of how SES factors may interact with investment choices, elevating the discussion surrounding the potential dark side implications of product recommendations on investor behaviors.

\begin{table}[ht!]
\footnotesize
  \caption{Hypothetical Excess Return Comparison Between Recommended and Non-Recommended Funds}
  \centering
    \begin{tabular}{cccc}
    \midrule
    \multicolumn{1}{c}{Investor Group} & \multicolumn{1}{c}{Non-Recommended Funds} &  \multicolumn{1}{c}{Recommended Funds} & \multicolumn{1}{c}{Difference} \\
    \midrule
    All Investors & -0.033 & -0.016 & 0.017** \\
    SES: Low & -0.036 & -0.010 & 0.026*** \\
    SES: Medium & -0.026 & -0.031 & -0.005 \\
    SES: High & -0.030 & -0.027 & -0.003 \\
    \midrule
    \multicolumn{4}{p{40em}}{\scriptsize Notes: {***} $p < 0.01$, {**} $p < 0.05$, {*} $p < 0.1$.  The significance level is based on an independent sample t-test.} \\
    \end{tabular}
  \label{tab:fundReturnCompareGroup}%
\end{table}%

\subsection{\label{fund reserversal}Fund Future Performance Reversal}
Besides investors' behavioral changes after following product recommendations, we also explore potential differences between recommended and non-recommended funds. Specifically, we examine the difference in future performance of recommended and non-recommended funds as potential explanations for worse performance when investors follow recommendations. Since our study intentionally narrows its focus to funds that marginally meet or miss the recommendation threshold, this design choice posits an intriguing question: Given that the annual returns of these recommended and non-recommended funds are closely aligned at the time of purchase, could their future performance diverge significantly? To explore this, we examine the correlation between current and future annual returns for both recommended and non-recommended funds in our data set.

Specifically, we calculate the mean value of each fund's current annual return and its annual return three months into the future. This three-month future return metric aligns well with the average holding period observed in our data set, which stands at approximately 3.3 months.

As revealed in Table~\ref{tab:fundFutureReturn}, recommended funds—those rank third, fourth, and fifth—exhibit higher annual returns than their non-recommended counterparts (funds rank sixth, seventh, and eighth) at the time of the recommendation. Intriguingly, this disparity narrows markedly after a three-month span. Put differently, on average, investors who opt for recommended funds experience less favorable returns after three months compared to those who select non-recommended funds (-0.230 vs. -0.144). This phenomenon of price reversals, which is also observed in broader financial markets, could be attributed to market overreactions or herd behavior, as highlighted in existing literature \citep{de1985does, atkins1990price, huang2010return}.

Consequently, apart from variations in investor behaviors, the observed reversal in future performance between recommended and non-recommended funds may offer another explanation as to why investors often fare worse when purchasing recommended funds. This insight provides another perspective on how fund recommendations negatively influence investment outcomes.

\begin{table}[ht!]
  \footnotesize
  \caption{Relation Between Current and Future Annual Return}
  \centering
    \begin{tabular}{llll}
    \midrule
    Variable & CurrentAnnualReturn & ThreeMonthAnnualReturn  & Gain\\
    \midrule
    Recommended Funds  & 0.569 & 0.339 & -0.230*** \\
    Non-Recommended Funds & 0.455 & 0.310  & -0.144*** \\
    Difference & 0.114*** & 0.029** &  \\
    \midrule
    \multicolumn{4}{p{40em}}{\scriptsize Notes: {***} $p < 0.01$, {**} $p < 0.05$, {*} $p < 0.1$. CurrentAnnualReturn is the current annual return of the fund. ThreeMonthAnnualReturn is the annual return of the fund after three months. Gain is the difference between ThreeMonthAnnualReturn and CurrentAnnualReturn. The significance level is based on Paired t-test.} 
    \\
    \end{tabular}
    \vspace{1em}
  \label{tab:fundFutureReturn}%
\end{table}%

\section{\label{conclusion}Conclusion}
As the number of retail investors trading on online investment platforms continues to soar, their increasing influences are propelling shifts in financial markets. These investors rely on online investment platforms, which adopt product recommendations to help investors find good financial products and realize higher investment returns. Whereas product recommendations can provide reliable quality signals for consumer goods using past ratings and sales, their quality signals appear unreliable for financial products when they rely on historical returns. In this paper, we examine the impact of product recommendations within the realm of financial products. We delve into how product recommendations affect investors' adoption, subsequent behavioral changes, and investment performance.

First, we reveal that investors follow product recommendations to make investment decisions even though recommendations based on historical returns cannot offer reliable quality signals. Further analyses reveal investors realize significantly worse investment returns when purchasing recommended funds (versus non-recommended ones). These findings suggest that platforms should be cautious when employing unreliable quality signals such as historical performance to design recommender systems. They should also clearly inform investors that historical returns cannot predict future returns and investors should not solely rely on product recommendations in their decisions. Platforms might also consider simultaneously providing recommendations based on other metrics such as fund risk and diversity so investors can obtain a more comprehensive picture of recommended funds. 

Second, our observations reveal a notable disparity: investors with lower SES exhibit a significantly higher tendency to follow product recommendations, in contrast to investors with medium and high SES. Alarmingly, low SES investors are the sole group experiencing significantly diminished investment returns after purchasing recommended funds. These findings raise concerns about the unintended consequences of product recommendations on investment platforms, as they appear to exacerbate wealth inequality. It is troubling that those who should benefit the most from financial protection, namely low SES investors, end up being the most adversely affected by these recommendations. In light of these findings, investment platforms should consider prioritizing the provision of educational resources to all investors, with a particular focus on those with low SES backgrounds. By equipping investors with fundamental knowledge of investments, platforms can empower them to make more informed decisions, better comprehend historical returns, and interpret financial data effectively, ultimately mitigating the negative impacts of product recommendations. Platforms can further enhance their commitment to safeguarding low SES investors by implementing additional protective measures. These measures may include offering guidance on diversifying investment portfolios and strategies for minimizing risks, thereby promoting a more secure and inclusive investment environment for all.

Third, our findings also reveal that investors significantly reduce search time when purchasing recommended funds versus non-recommended ones. This reduction in research time can result in less-informed investment decisions, a trend particularly pronounced among low SES investors, who also heavily rely on product recommendations. To address this issue and encourage more informed decision-making, investment platforms could consider implementing gamification elements that motivate investors to dedicate more effort. This might involve incentivizing investors to explore a certain number of funds or different fund types before finalizing their purchases. Fourth, our simulation analyses indicate that investors who opt for recommended funds often redeem them at less opportune times compared to those who invest in non-recommended funds. While platforms cannot offer specific guidance on the perfect redemption timing, they can enhance the investor experience by furnishing more comprehensive data on historical trends and market cycles. This additional information can assist investors in making more considered and judicious redemption decisions. 

Finally, our findings also yield crucial implications for policymakers concerning the utilization of recommendations within financial platforms. Even in the case of non-personalized recommendations, without appropriate design, they may still yield adverse and unintended consequences for investors. Most significantly, this could potentially give rise to systemic risks within the financial market, as numerous investors may make similar investment choices by following these recommendations, leading to concurrent losses that could propagate throughout the financial system \citep{benoit2017risks}. To avert such detrimental outcomes for investors and the broader financial market, investment platforms, in collaboration with policymakers, should devise effective strategies for monitoring the deployment of product recommendations. Furthermore, they should advocate best practices to support the decision-making processes of investors, thereby enhancing the overall resilience and stability of the financial ecosystem.

We have sought to pinpoint the specific impact of product recommendations within the context of financial products, using the example of mutual funds. As an avenue for future research, we advocate expanding the scope to include an array of other prevalent financial instruments like futures, options, and insurance products. Our study is chiefly concerned with the mechanism of product recommendations—a design ubiquitously employed across e-commerce and financial platforms alike. Moving forward, it would be illuminating to examine the influence of other common platform designs such as portfolio comparison or custom portfolios. By broadening the range of financial products and platform designs studied, future research could offer a more comprehensive understanding of how various platforms shape investor behaviors and financial outcomes.

\section{Funding and Competing Interests}

All authors certify that they have no affiliations with or involvement in any organization or entity with any financial or non-financial interest in the subject matter or materials discussed in this manuscript. The authors have no funding to report.

\bibliographystyle{informs2014}
\bibliography{reference}

\newpage
\begin{appendices}

\section{\label{A} Investor Characteristics}
\setcounter{table}{0}
\renewcommand{\thetable}{A\arabic{table}}

\begin{table}[ht!]
\footnotesize
 \caption{Summary Statistics of Investor Characteristics}
  \centering
    \begin{tabular}{ccccc}
    \midrule
    \multicolumn{1}{c}{Characteristics} & \multicolumn{1}{c}{Description}  & \multicolumn{1}{c}{\# of Investors} & \multicolumn{1}{c}{Value} & \multicolumn{1}{c}{Percentage} \\
    \midrule
    \multirow{3}[1]{*}{Income} & Not Stable & 9,328 & 0 & 74.98\% \\
          & Stable & 2,125 & 1 & 17.08\% \\
          & Stable and High & 988 & 2 & 7.94\% \\
    \midrule
    \multirow{3}[0]{*}{Liability} & High & 246 & 0 & 1.98\% \\
          & Medium & 10,531 & 1 & 84.65\% \\
          & Low & 1,664 & 2 & 13.38\% \\
    \midrule
    \multirow{3}[0]{*}{Education} & High School or Below & 9,696 & 0 & 77.94\% \\
          & Bachelor & 2,223 & 1 & 17.87\% \\
          & Graduate & 522 & 2 & 4.20\% \\
    \midrule
    \multicolumn{5}{p{36em}}{\scriptsize Notes: Investor characteristics are encoded into ordinal values for investor segmentation by socioeconomic status. Higher values represent higher income, education, and fewer liabilities.} \\
    \end{tabular}
  \label{tab:investorSummary}
\end{table}

\newpage
\section{\label{B} Robustness Checks}
\setcounter{table}{0}
\renewcommand{\thetable}{B\arabic{table}}
\begin{table}[ht!]
\caption{Robustness Check: Falsification Exercise \& Alternative Bandwidth}
\footnotesize
  \centering
\begin{tabular}{lcccc}
    \midrule
    \multicolumn{1}{l}{Dependent Variable:} & \multicolumn{1}{c}{(1)} & \multicolumn{1}{c}{(2)} & \multicolumn{1}{c}{(3)} & \multicolumn{1}{c}{(4)} \\
    \multicolumn{1}{l}{log(NumberOfPurchases)} & \multicolumn{1}{c}{} & \multicolumn{1}{c}{}  & \multicolumn{1}{c}{}\\
    \midrule
    \multicolumn{1}{l}{Recommended} & \multicolumn{1}{c}{-0.012} & \multicolumn{1}{c}{0.008} & \multicolumn{1}{c}{0.029***} & \multicolumn{1}{c}{0.028***} \\
    \multicolumn{1}{l}{} & \multicolumn{1}{c}{(0.011)} & \multicolumn{1}{c}{(0.010)} & \multicolumn{1}{c}{(0.010)} & \multicolumn{1}{c}{(0.007)} \\
    \multicolumn{1}{l}{Fund Type Dummies} & \multicolumn{1}{c}{Yes} & \multicolumn{1}{c}{Yes} & \multicolumn{1}{c}{Yes} & \multicolumn{1}{c}{Yes} \\
    \multicolumn{1}{l}{Month Dummies} & \multicolumn{1}{c}{Yes} & \multicolumn{1}{c}{Yes} & \multicolumn{1}{c}{Yes} & \multicolumn{1}{c}{Yes} \\
    \multicolumn{1}{l}{Day of Week Dummies} & \multicolumn{1}{c}{Yes} & \multicolumn{1}{c}{Yes} & \multicolumn{1}{c}{Yes} & \multicolumn{1}{c}{Yes} \\
    \multicolumn{1}{l}{Control Variables} & \multicolumn{1}{c}{Yes} & \multicolumn{1}{c}{Yes} & \multicolumn{1}{c}{Yes} & \multicolumn{1}{c}{Yes} \\
    \multicolumn{1}{l}{Number of Observations} & \multicolumn{1}{c}{7,278} & \multicolumn{1}{c}{7,278} & \multicolumn{1}{c}{7,278} & \multicolumn{1}{c}{14,556} \\
    \multicolumn{1}{l}{\textcolor[rgb]{ .133,  .133,  .133}{$R^2$}} & \multicolumn{1}{c}{0.173} & \multicolumn{1}{c}{0.182} & \multicolumn{1}{c}{0.197} & \multicolumn{1}{c}{0.171} \\
    \midrule
    \multicolumn{5}{p{30em}}{\scriptsize Notes: {***} $p<0.01$, {**} $p<0.05$, {*} $p<0.1$. The dependent variable in Columns (1) - (4) is log(NumberOfPurchases). Control variables include management fees, purchase fees, custodial fees, asset size, dummy for valuable fund brand, fund past returns, fund risk level, and fund tenure. Columns (1) and (2) show the results of the falsification exercise. Column (1) compares funds in fourth and fifth places (both are recommended). Column (2) compares funds in the sixth and seventh places (both are non-recommended). Column (3) uses a smaller bandwidth that includes only funds in the fifth (recommended) and sixth (non-recommended) places in the sample. Column (4) uses a smaller bandwidth that includes only funds in the fourth to seventh places in the sample. Robust standard errors are in parentheses.}\\
    \end{tabular}
  \label{tab:falsificationAlternateBandwidth}
\end{table}

\begin{table}[ht!]
\footnotesize
\caption{Robustness Check: Alternative Measurements}
  \centering
    \begin{tabular}{p{8em}lccc}
    \midrule
    \multicolumn{1}{l}{Dependent Variable:} & \multicolumn{1}{c}{(1)} & \multicolumn{1}{c}{(2)} & \multicolumn{1}{c}{(3)} \\
    \multicolumn{1}{l}{} & \multicolumn{3}{c}{}\\
    \midrule
    \multicolumn{1}{l}{Recommended} & \multicolumn{1}{c}{0.052***} & \multicolumn{1}{c}{0.163***} & \multicolumn{1}{c}{20.440***} \\
    \multicolumn{1}{r}{} & \multicolumn{1}{c}{(0.006)} & \multicolumn{1}{c}{(0.018)} & \multicolumn{1}{c}{(2.787)} \\
     \multicolumn{1}{l}{Fund Type Dummies} & \multicolumn{1}{c}{Yes} & \multicolumn{1}{c}{Yes} & \multicolumn{1}{c}{Yes} \\
    \multicolumn{1}{l}{Month Dummies} & \multicolumn{1}{c}{Yes} & \multicolumn{1}{c}{Yes} & \multicolumn{1}{c}{Yes} \\
    \multicolumn{1}{l}{Day of Week Dummies} & \multicolumn{1}{c}{Yes} & \multicolumn{1}{c}{Yes} & \multicolumn{1}{c}{Yes} \\
    \multicolumn{1}{l}{Control Variables} & \multicolumn{1}{c}{Yes} & \multicolumn{1}{c}{Yes} & \multicolumn{1}{c}{Yes} \\
    \multicolumn{1}{l}{Number of Observations} & \multicolumn{1}{c}{21,834} & \multicolumn{1}{c}{21,834} & \multicolumn{1}{c}{21,834} \\
    \multicolumn{1}{l}{\textcolor[rgb]{ .133,  .133,  .133}{$R^2$}} & \multicolumn{1}{c}{0.156} & \multicolumn{1}{c}{0.144} & \multicolumn{1}{c}{0.066} \\
    \midrule
    \multicolumn{4}{p{28em}}{\scriptsize Notes: {***} $p<0.01$, {**} $p<0.05$, {*} $p<0.1$. The dependent variable in Column (1) is log(NumberOfInvestors), log(PurchaseAmount) in Column (2), and MoneyInflow in Column (3). Columns (1) - (3) use alternative measurements to operationalize the effect of recommendation on purchase behavior. Control variables include management fees, purchase fees, custodial fees, asset size, dummy for valuable fund brand, fund past returns, fund risk level, and fund tenure.  Robust standard errors are in parentheses.} \\
    \end{tabular}
  \label{tab:altDV}
\end{table}

\begin{table}[ht!]
\footnotesize
\caption{Robustness Check: Alternative Model Specification}
  \centering
    \begin{tabular}{lc}
    \midrule
    \multicolumn{1}{l}{Dependent Variable:} & \multicolumn{1}{c}{(1)} \\
    \multicolumn{1}{l}{NumOfPurchases} & \multicolumn{1}{c}{} \\
    \midrule
    \multicolumn{1}{l}{Recommended} & \multicolumn{1}{c}{0.401**} \\
    \multicolumn{1}{r}{} & \multicolumn{1}{c}{(0.061)} \\
    \multicolumn{1}{l}{Fund Type Dummies} & \multicolumn{1}{c}{Yes} \\
    \multicolumn{1}{l}{Month Dummies} & \multicolumn{1}{c}{Yes} \\
    \multicolumn{1}{l}{Day of Week Dummies} & \multicolumn{1}{c}{Yes} \\
    \multicolumn{1}{l}{Control Variables} & \multicolumn{1}{c}{Yes} \\
    \multicolumn{1}{l}{Number of Observations} & \multicolumn{1}{c}{21,834} \\
    \multicolumn{1}{l}{\textcolor[rgb]{ .133,  .133,  .133}{Pseudo $R^2$}} & \multicolumn{1}{c}{0.302} \\
    \midrule
    \multicolumn{2}{p{26em}}{\scriptsize Notes: {***} $p<0.01$, {**} $p<0.05$, {*} $p<0.1$. NumberOfPurchases is the dependent variable in Column (1). Column (1) displays the result using a Poisson regression model. Control variables include management fees, purchase fees, custodial fees, asset size, dummy for valuable fund brand, fund past returns, fund risk level, and fund tenure.  Robust standard errors are in parentheses.} \\
    \end{tabular}
  \label{tab:altModelSubsample}
\end{table}

\end{appendices}

\end{document}